\documentclass[aps,prb,twocolumn,groupedaddress,showpacs,floatfix]{revtex4}
\usepackage{graphicx}
\usepackage{amsgen,amsfonts,amsbsy,amssymb,amsmath}
\begin{document}

\title{One-electron spectra and susceptibilities of 3D electron gas from self-consistent solutions of Hedin's equations}
\author{A.L. Kutepov\footnote{e-mail: kutepov@physics.rutgers.edu}, and G. Kotliar}

\affiliation{Department of Physics and Astronomy, Rutgers
University, Piscataway, NJ 08856}

\begin{abstract}
A few approximate schemes to solve the Hedin equations \cite{pr_139_A796} self-consistently introduced in (Phys. Rev. B 94, 155101 (2016)) are explored and tested for the 3D electron gas at metallic densities. We calculate one electron  spectra, dielectric properties, compressibility, and correlation energy. Considerable reduction in the calculated band width (as compared to the self consistent GW result) has been found when vertex correction was used for both polarizability and self energy. Generally, it is advantageous to obtain the diagrammatic representation of polarizability from the definition of this quantity as a functional derivative of the electronic density with respect to the total field (external plus induced). For self energy, the first order vertex correction seems to be sufficient for the range of densities considered. Whenever it is possible, we compare the accuracy of our vertex-corrected schemes with the accuracy of the self-consistent quasi-particle GW approximation (QSGW), which is less expensive computationally. We show that QSGW approach performs poorly and we relate this poor performance with an inaccurate description of the screening in the QSGW method (with an error comprising a factor 2-3 in the physically important range of momenta).
\end{abstract}

 \maketitle

\section*{Introduction}
\label{intro}

Many body perturbation theory (MBPT) diagrammatic approaches offer a path to the solution of the quantum many body problem of solids, which
complements alternative methods such as QMC (Quantum Monte Carlo) and CC (Couple Cluster). This approach received attention for over half a century but there is no complete
understanding of how the different selection of diagrams performs for different physical quantities.  These insights are  important in  the search for
predictive first principles methods for correlated solids. In this work we investigate these questions in the framework of  the homogeneous
electron gas (HEG) in a neutralizing positively charged background. The HEG is very useful  for testing, in a simplified setting, the methods
presently being developed to study the electronic structure of  solids for two reasons: it requires less computational effort and there is a
natural benchmark  since some properties  have also been  calculated using   QMC methods.
\cite{prl_45_566,prb_79_085414,prb_87_045131,prb_50_14838,prb_45_13244} This model describes the properties of alkali metals well.

Most common uses of diagrammatic approaches are based on non-interacting Green functions such as LDA (Local Density Approximation) Green's function. In this work, however, we are interested in self consistent (sc) diagrammatic approaches. Hence    we do not consider  ambiguities related to the
choice of non interacting Green's function for reference.
We consider two classes of methods. One, initiated by Hedin,\cite{pr_139_A796}, which   carries out a perturbative expansion in the fully self consistent
(renormalized) Green's function (which obeys  the Dyson equation) and screened interaction W. An alternative philosophy (QSGW) uses the lowest
order diagrams for polarizability and self energy with Green's function which is determined by means of quasi-particle self
consistency condition.\cite{prb_76_165106} In this work we perform QSGW calculations using a previously introduced linearized approach\cite{prb_85_155129} that has an advantage of being implementable on Matsubara frequencies.

An  important cornerstone in the application of self-consistent MBPT-based approximations to the HEG, is the work by Holm and von Barth,\cite{prb_57_2108}
where the authors applied scGW  approximation  (i.e. lowest order diagram in the perturbation series in terms of G and W) to calculate the total
energy and spectra of HEG. Their principal conclusions are that scGW severely overestimates the band width but gives the total energy very
close to the QMC results. Thus, the work [\onlinecite{prb_57_2108}] raised the question of  how (if at all) one can get accurate spectra of HEG using
scMBPT approach.   Progress on this question was made by Shirley \cite{prb_54_7758} who showed that if an accurate W is known (he used QMC input
to evaluate W) then the self-consistency in Green's function G and the lowest order vertex corrections in self energy mostly cancel out each
other. This observation justifies, in a certain degree, the so called one-shot $G_{0}W$ approaches ( lowest order approach using  a non interacting Greens function
$G_{0}$) without vertex corrections to self energy. Takada \cite{prl_87_226402,prb_84_245134} used QMC data to parametrize electron-hole four-point
irreducible interaction and performed sc vertex-corrected GW calculations with the vertex defined from a condition that Ward Identity (WI) is
satisfied. In this respect, his approach resembles the idea of QSGW method, where WI is imposed by construction. The importance of vertex corrections in electron gas studies was reported recently in non-self-consistent calculations.\cite{prb_93_235113,prl_117_206402} From other studies on the subject, one can mention interesting applications of quantum chemical methods (first of all of ab-initio coupled-cluster theory) to studying the spectra\cite{prb_93_235139} and correlation energy\cite{prl_110_226401,prl_112_133002} of electron gas.

In this study  we go beyond the earlier diagrammatically inspired works by removing an important limitation related to the use of  Monte-Carlo data for parametrization of screened interaction or electron-hole four-point irreducible interaction. We examine fully self-consistent (in both G
and W) diagrammatic schemes involving diagrams of higher order. We study relative importance of different diagrams for polarizability and self energy.

The paper begins with a brief presentation of self-consistent schemes we use to solve Hedin's equations (section \ref{meth}). Section \ref{res} provides the
results obtained and a discussion. The conclusions are given afterwards.

\section{Method}
\label{meth}

Detailed account of the vertex-corrected schemes we use in this work has been given in Ref.[\onlinecite{prb_94_155101}]. For completeness, below, we briefly repeat the essentials of the approach and point out the simplifications in technical implementation in the case of electron gas.

We solve Hedin's equations \cite{pr_139_A796} self-consistently using different approximations for three point vertex function $\Gamma$.
Three point vertex function enters formally exact expressions for polarizability and self energy (in space-time variables)

\begin{align}\label{vrt1}
P(12)=\sum_{\alpha}G^{\alpha}(13)\Gamma^{\alpha}(342)G^{\alpha}(41),
\end{align}
\begin{align}\label{vrt2}
\Sigma^{\alpha}(12)=-G^{\alpha}(13)\Gamma^{\alpha}(324)W(41),
\end{align}
where the integration/summation over repeated arguments is understood, and $\alpha$ is the spin index.

We consider two different types of approximations for $\Gamma$. The first type consists in expanding vertex function in terms
of the screened interaction to a specified order. Keeping only zero order term ($\Gamma=1$) in both $P$ and $\Sigma$ corresponds to the famous GW
approximation. We will also consider expansion of the vertex up to the first order ($\Gamma_{1}=1+WGG$) in both polarizability and self energy expressions. This approximation is conserving (like GW) as the corresponding P and $\Sigma$ can alternatively be obtained by
differentiating the same $\Psi$-functional.\cite{ijmpb_13_535,prb_63_115110}

\begin{figure}[b]
\includegraphics[width=7.0 cm]{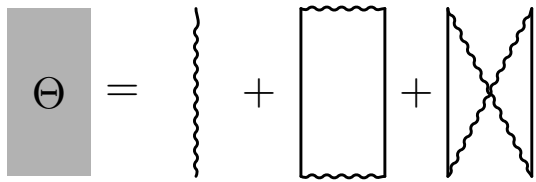}
\caption{The approximation for the irreducible 4-point vertex function $\Theta$.} \label{theta_appr}
\end{figure}

The second type of approximation for $\Gamma$ consists in solving the Bethe-Salpeter equation
\begin{align}\label{Vert_0}
\Gamma^{\alpha}(123)&=\delta(12)\delta(13)\nonumber\\&+\frac{\delta
\Sigma^{\alpha}(12)}{\delta
G^{\beta}(45)}G^{\beta}(46)\Gamma^{\beta}(673) G^{\beta}(75),
\end{align}
with a certain approximate expression for the functional derivative $\Theta=\frac{\delta \Sigma}{\delta G}$ in (\ref{Vert_0}). We will consider
two expressions for the kernel $\Theta$ in this work. The first is obtained by using the GW form for $\Sigma$ in the functional derivative and
neglecting the derivative of the screened interaction $\frac{\delta W}{\delta G}$, i.e. $\Theta=W$ (we will call the corresponding vertex as
$\Gamma^{0}_{GW}$). Diagrammatically it corresponds to keeping only the first term on the right hand side of Fig.\ref{theta_appr}. In the second
approximate expression for $\Theta$ we also are using GW form for self energy in the functional derivative but we keep the terms up to the
second order in $W$ in the derivative $\frac{\delta W}{\delta G}$ (we will use abbreviation $\Gamma_{GW}$ for the corresponding vertex). In
this case, the obtained vertex function corresponds to keeping all three terms for $\Theta$ (Fig.\ref{theta_appr}). It is important to point out
that the diagrams resulted from $\frac{\delta W}{\delta G}$ allow the spin flips (as it is clear from Fig.\ref{theta_appr}), the importance of
which was pointed out in Ref.[\onlinecite{prb_71_045323}].

In the particular case, when G and W have been found self-consistently with $\Sigma=GW$ and
$P=GG$, vertex $\Gamma_{GW}$ yields physical polarizability in scGW approximation
(defined as a functional derivative of electronic density with respect to the total electric field).
In other cases, when $\Gamma_{GW}$ is evaluated with $\Sigma$ and/or $P$ including additional diagrams, the
kernel shown in Fig. \ref{theta_appr} is only an approximation to the derivative $\frac{\delta \Sigma}{\delta G}$, and as a result,
the vertex $\Gamma_{GW}$ doesn't provide physical $P$ anymore. Thus, in a search for an optimal approximation, we have to trade between the number of diagrams included in $\Sigma$ in the Dyson equation for $G$ and the degree of the "deviation" of polarizability from the physical one. Another potential problem which can arise when higher order diagrams are summed up uncontrollably is an appearance of negative spectral functions. This issue has been known since the works by Minnhagen\cite{jpc_7_3013,jpc_8_1535} and a solution (positive-definite diagrammatic expansion for the spectral function and for the density-response spectrum) was found recently in Refs.[\onlinecite{prb_90_115134,prb_91_115104,prl_117_206402}]. Below, we demonstrate that our calculated spectral functions are positive.

Thus, in a search for optimal approximation, we have to trade between the
accuracy of $\Sigma$ in the Dyson equation for $G$ and the degree of the "deviation" of polarizability from the physical one. Another potential problem which can arise when higher order diagrams are summed up uncontrollably is the appearance of negative spectral functions. This issue has been known since the works by Minnhagen {J.Phys.C 7, 3013(1974), J.Phys.C 8,1535(1975)} and a solution (positive-definite diagrammatic expansion for the spectral function and for the density-response spectrum) was found recently in Refs.[PRB 90,115134(2014), PRB 91,115104(2015), PRL 117,206402(2016)]. Below, we demonstrate the positivity of our calculated spectral functions.

\begin{table}[b]
\caption{Diagrammatic representations of polarizability and self energy in sc schemes of solving the Hedin equations. Arguments in square
brackets specify $G$ and $W$ which are used to evaluate the vertex function. Other details are explained in the main text.} \label{sc_schemes}
\begin{center}
\begin{tabular}{@{}c c c} Scheme  & $P$ & $\Sigma$\\
\hline
A & $GG$ & $GW$ \\
B & $G\Gamma_{1}[G;W]G$ & $G\Gamma_{1}[G;W]W$ \\
C & $\underline{G}\Gamma_{GW}[\underline{G};\underline{W}]\underline{G}$ & $G\overline{W}$ \\
D & $\underline{G}\Gamma_{GW}[\underline{G};\underline{W}]\underline{G}$ & $G\Gamma_{1}[G;\overline{W}]\overline{W}$ \\
E & $G\Gamma_{GW}[G;W]G$ & $G\Gamma_{1}[G;W]W$ \\
G & $G\Gamma^{0}_{GW}[G;W]G$ & $G\Gamma_{1}[G;W]W$
\end{tabular}
\end{center}
\end{table}

With our four (as specified above) approximations for the vertex functions ($1;\Gamma_{1};\Gamma^{0}_{GW};\Gamma_{GW}$) we are able to form
different self-consistent schemes for solving Hedin's equations by selecting the vertex to be used in polarizability (\ref{vrt1}) and the
vertex to be used in self energy (\ref{vrt2}). As all our vertices are approximate, they don't have to be the same in $P$ and in $\Sigma$. We have tried different combinations and below we will show the results obtained with a reasonable subset of them. To
distinguish the approaches we will use the same notations as the ones introduced in the Ref.[\onlinecite{prb_94_155101}]. For the convenience, we have collected them in Table \ref{sc_schemes} (slightly modified Table I from the Ref.[\onlinecite{prb_94_155101}]) and we repeat here their definitions. Scheme A is the scGW approach.  It is conserving in Baym-Kadanoff definition,\cite{pr_124_287} but generally its accuracy is poor when one considers spectral properties of solids.\cite{prb_80_041103,prb_85_155129,prl_89_126402} Another
conserving sc scheme is scheme B. It uses the same first order vertex $\Gamma_{1}$ in both $P$ and $\Sigma$. Scheme C is based on the "physical" polarizability (preserves charge microscopically). In scheme C, we perform the scGW calculation first. Underlined $\underline{G}$ and $\underline{W}$ in Table \ref{sc_schemes}
mean that the corresponding quantities are taken from the scGW run. Then the vertex $\Gamma_{GW}[\underline{G};\underline{W}]$ is evaluated and it is
used to calculate polarizability and corresponding screened interaction $\overline{W}$. We use a bar above the $W$ to indicate that this
quantity is evaluated using $\underline{G}$ and $\underline{W}$ from the scGW calculation, but it is not equal to $\underline{W}$ because it includes
vertex corrections through the polarizability. This $\overline{W}$ is fixed (in scheme C) during the following iterations where only the self
energy $\Sigma=G\overline{W}$ and $G$ are renewed. So, scheme C doesn't include the vertex in $\Sigma$ explicitly but only through $\overline{W}$.
Scheme D is similar to scheme C. It also is based on physical polarizability but it uses the first order vertex in self energy
explicitly (skeleton diagram). In scheme D the screened interaction $\overline{W}$ is fixed at the same level as in scheme C, but the final iterations involve the
renewal of not only $G$ and $\Sigma$, but also $\Gamma_{1}$. Scheme E is fully self-consistent (both $G$ and $W$ are renewed on every iteration till
the end). Scheme E doesn't preserve the charge exactly and can be considered as a result of a trade between the accuracy of self energy and the degree of deviation of polarizability from the physical one. Scheme G is similar to scheme E, but with a simplified
Bethe-Salpeter equation for the corresponding vertex $\Gamma^{0}_{GW}$ (the diagrams with spin-flips are neglected in the kernel of the Bethe-Salpeter
equation).

In accordance with the arguments above, we have found that schemes with vertex $\Gamma_{GW}$ in $P$ and with vertex of increasing ($>1$)
order in $\Sigma$ (scheme F in the Ref.[\onlinecite{prb_94_155101}]) result in non-physical polarizability (first of all in its improper $\mathbf{q}\rightarrow 0$ behavior) and in the deterioration
of the accuracy in calculated properties. We will not consider them further in this work.

For 3D electron gas, we solve Hedin's equations in a periodic
cubic box with equidistant $54\times54\times54$ mesh. The box contains 729 electrons. We use Matsubara's formalism with electronic temperature 1000K. We
do not use plasmon pole approximation and we treat full frequency
dependence of W, as opposite to the often use of its zero frequency
limit when solving Bethe-Salpeter equation for insulators
\cite{prb_92_035202} or, in the recent paper on the electron gas.\cite{prb_93_235113} Detailed formulas, presented in the Appendix of the Ref.[\onlinecite{prb_94_155101}] are simplified for the electron gas considerably by omitting the indexes associated with the band states, the muffin-tin orbitals, and the product basis.

\section{Results}
\label{res}

\begin{table}[t]
\caption{Band widths (eV) of the 3D electron gas compared with the
results from Ref.[\onlinecite{prb_54_7758}], where the QMC input was
partially used.} \label{bw_comp}
\begin{center}
\begin{tabular}{@{}c c c c c} $r_{s}$  & 2 & 3 & 4 & 5\\
QSGW &13.48 & 5.75 & 3.10 & 1.92 \\
A &13.61 & 6.08 & 3.44 & 2.21 \\
B &12.53 & 5.51 & 3.07 & 1.94 \\
C &13.21 & 5.90 & 3.28 & 2.10 \\
D &11.54 & 5.22 & 2.85 & 1.79 \\
E &11.59 & 5.10 & 2.78 & 1.73 \\
G &11.79 & 5.20 & 2.86 & 1.80 \\
\hline
[\onlinecite{prb_54_7758}] & 11.57(5) & 5.04(4) & 2.66(4) & 1.72(4)
\end{tabular}
\end{center}
\end{table}

\begin{figure}[b]
\centering
\includegraphics[width=8.0 cm]{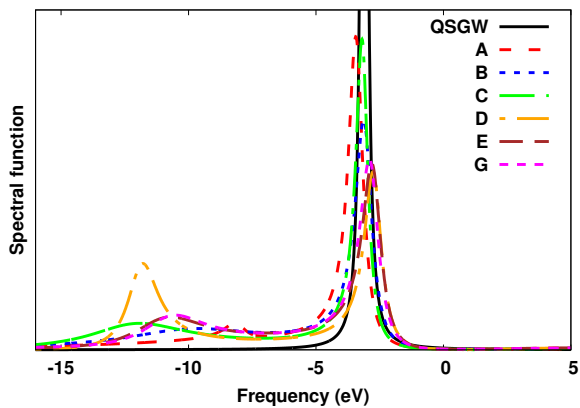}
\caption{(Color online) Spectral function ($\mathbf{k}=0$, arb. units) of the electron gas for $r_{s}=4$.} \label{sf4}
\end{figure}

In Table \ref{bw_comp} we compare our results for band width with those obtained by Shirley,\cite{prb_54_7758} who based the calculations
(partially) on QMC input. Band width was determined as a difference between the pole in the spectral function at \textbf{k}=0 and chemical potential. If we assume that Shirley's results are close to the exact ones, we can draw certain conclusions about our approaches. As one can conclude from the Table, three schemes (D, E, and G) show the best performance with small differences between themselves. Common for these three schemes are two facts: they all include a solving of the Bethe-Salpeter equation for polarizability (but slightly differently as it was explained above) and they all apply first order vertex correction in self energy. Scheme C also involves solving of BSE for polarizability, but it doesn't use vertex correction to self energy and, as a result, shows worse performance. Similarly, conserving scheme B, which applies first order vertex corrections to the $P$ and $\Sigma$, shows worse performance, because it misses the effects of BSE in W. Nevertheless, scheme B seems to be better than scheme C, demonstrating the importance of vertex corrections in self energy. Performance of the QSGW approach is slightly better than performance of scGW, but it is not competitive with schemes D, E, or G. Thus, for the studied range of densities of electron gas, QSGW cannot be considered as a reasonable approximation in terms of its predictive power.

\begin{figure}[t]
\centering
\includegraphics[width=8.0 cm]{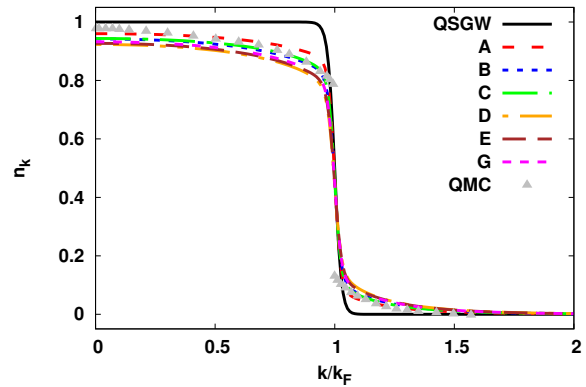}
\caption{(Color online) Electron occupations in the electron gas for $r_{s}=4$. QMC data are from Ref.[\onlinecite{prl_107_110402}].} \label{n_k}
\end{figure}

An example of $\mathbf{k}$-resolved spectral functions is shown in Fig.\ref{sf4}. First, we would like to point out that all our calculated spectral functions are positive (they are also positive for non-zero momenta). As one can see there is a well defined quasi-particle peak near -3 eV. All approaches (excluding QSGW) also show plasmon satellite at roughly -11 eV with its exact position being sensitive to the specific scheme used. The positions of plasmon satellites are also sensitive to the quality of analytical continuation which we performed using the method of Vidberg and Serene.\cite{jltp_29_179} We checked the accuracy of this method to be rather good to determine the positions of quasi-particle peaks, but we would give an error bar about 1 eV for the positions of plasmon satellite peaks. The accuracy in the calculated positions of plasmon satellite peaks can be improved using more points in frequency summations and time integrations when evaluating the higher order diagrams with, however, corresponding increase in the computation time. Plasmon satellites in electron gas have been studied recently using the GW+Cumulant approach.\cite{prb_93_235446,epjb_89_238,prl_117_206402}

Figure \ref{n_k} presents electron occupations $n_{k}$ obtained for $r_{s}=4$. Temperature effects are responsible for a slight deviation of the QSGW curve from the perfect step function. Other approaches also have (in addition to the temperature effects) a correlation-related spectral weight transfer. We compare our results with available QMC data.\cite{prl_107_110402} However, it is hard to make this comparison conclusive. First of all, the mentioned above temperature effects make our calculated momentum distribution smoother than it would be at T=0K. Second, QMC data are essentially based on the extrapolation to the thermodynamic limit (the inset in Fig.1 of Ref.[\onlinecite{prl_107_110402}] shows that the shape of the QMC curve is almost altogether the result of an extrapolation). Nevertheless, one can point out that at \textbf{k}=0 all our schemes (excluding QSGW) show smaller values of $n_{k}$ than QMC. Close to the Fermi momentum, our vertex corrected schemes seem to be closer to the QMC data than scGW result. But it is hardly possible to say which scheme is the best in terms of this physical quantity.

\begin{figure}[b]
\centering
\includegraphics[width=8.0 cm]{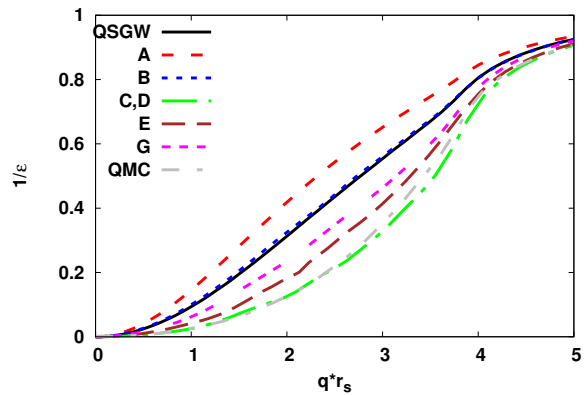}
\caption{(Color online) Inverse static dielectric function of 3D
electron gas, $r_{s}=4$. We use a fit to the QMC data as provided in Fig.6 of the Ref.[\onlinecite{prb_50_14838}] for comparison. Schemes C and D give identical results for this quantity.} \label{e1_4}
\end{figure}

In figure \ref{e1_4} the static (zero-frequency) inverse dielectric
function for 3D electron gas is shown for $r_{s}=4$. As one can see
from the graph, we are able to improve the agreement with QMC data
considerably when using the vertices of increased complexity. Clearly, the best dielectric function is obtained from the "physical" polarizability (schemes C and D), even if the last corresponds to the scGW approximation (in the sense that the diagrams are evaluated using G and W from scGW). One can also point out the importance of including
the diagrams with spin flips (scheme E results in better dielectric function than scheme G does). At the same time scheme E is worse than schemes C/D which reflects the above mentioned fact about trading - additional diagram in self energy in scheme E violates the requirement of polarizability to be physical. One can also relate the shortcomings of QSGW approach in the one-electron spectra to the poor description of the screening. As one can see from figure \ref{e1_4}, in the physically important range ($q*r_{s}$=0.5-3.0) the inverse dielectric function in QSGW approximation is larger than the one from QMC data by a factor 2-3.

\begin{figure}[t]
\centering
\includegraphics[width=8.0 cm]{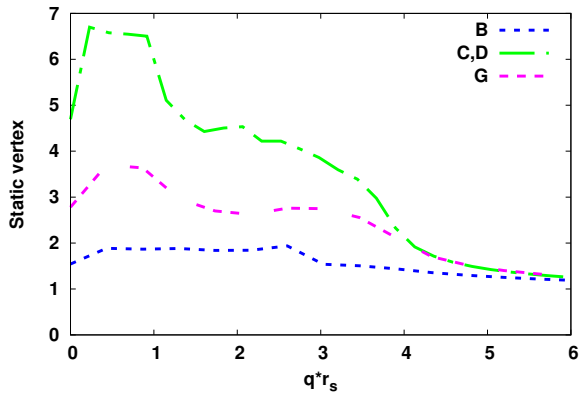}
\caption{(Color online) Static vertex ($\nu=0$) as a function of bosonic momentum \textbf{q} for $r_{s}=4$. Fermionic frequency and momentum correspond to their values at the Fermi surface. Schemes C and D give identical results for this quantity.} \label{vrt_stat}
\end{figure}

\begin{figure}[b]
\centering
\includegraphics[width=8.0 cm]{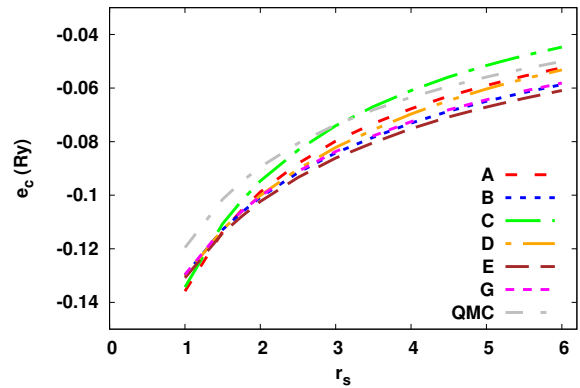}
\caption{(Color online) Correlation energy of 3D electron gas as obtained from conserving approximations. The QMC results are cited from the Ref.[\onlinecite{prl_45_566}].} \label{e_c}
\end{figure}

A certain insight on the origin of differences in the dielectric function obtained with approximate methods can be gained when one looks at the static vertex as a function of bosonic momentum (Fig. \ref{vrt_stat}). It is clear that the range of momenta where the calculated dielectric function shows the largest differences ($q*r_{s}=0-4$) correlates very well with the range of momenta where the humps in the vertex function show very different heights.

In Fig.\ref{e_c} we present the calculated correlation energy of the electron gas as a function of $r_{s}$. It was obtained as the difference between the expectation value of the Hamiltonian corresponding to the selected level of approximation and the expectation value of the Hamiltonian in the Hartree-Fock approximation. In all vertex-corrected schemes the exchange-correlation part was evaluated as a convolution of Green's function and self energy. Excluding scheme C which misses vertex corrections to self energy and, as a result, shows rather different from other schemes behavior, one can conclude that vertex corrections make the correlation energy more negative as compared to the correlation energy obtained in scGW approach. Only at $r_{s}<1.5$ this tendency is reversed. Thus, if we assume that available QMC data are exact, we have to state that vertex corrections systematically worsen the scGW result. However, the slope of the curves seems to get better at least in some of the vertex corrected schemes. As one can see, the deviation from QMC data in fully self-consistent schemes B, E, and G is almost $r_{s}$-independent.

\begin{table}[t]
\caption{Correlation energy (Ry) of 3D electron gas in scGW approximation compared with the
QMC data.} \label{et_gw}
\begin{center}
\begin{tabular}{@{}c c c c c} Method $\setminus r_{s}$  & 1 & 2 & 4 & 5\\
scGW [\onlinecite{prl_83_788}] & & -0.0901 &  -0.064 & \\
scGW [\onlinecite{prb_63_075112}] &-0.1156 & -0.0872 & -0.061 & -0.0538 \\
scGW [\onlinecite{prb_95_195131}] &-0.137 & -0.0996 & -0.0686 & -0.060 \\
scGW, present work &-0.1358 & -0.0988 & -0.0677 & -0.0591 \\
\hline
QMC [\onlinecite{prl_45_566}] & -0.1196 & -.0902 & -0.0638 & -0.0562
\end{tabular}
\end{center}
\end{table}

One more point related to the issue of correlation energy is its precise value obtained in scGW approximation. Since its first evaluation by Holm and von Barth\cite{prb_57_2108}, the consensus was that scGW approximation gives very accurate total energies of three-dimensional electron gas. However, our calculated energies (in scGW) are systematically more negative than the ones reported in earlier papers and in QMC studies. To make this point clearer we present the numbers in Table \ref{et_gw}. The most recent publication by Van Houcke et al.\cite{prb_95_195131} agrees well with our results. Our data and the data from Ref.[\onlinecite{prb_95_195131}] are almost identical with a discrepancy of about 0.001 Ry or less. Thus, common belief in reliability of scGW total energies for electron gas needs to be reconsidered.

\begin{figure}[b]
\centering
\includegraphics[width=8.0 cm]{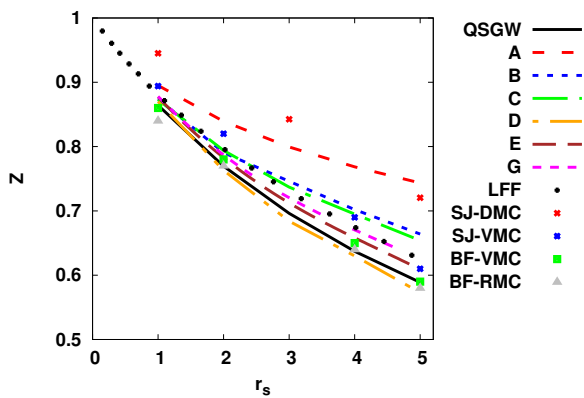}
\caption{(Color online) Quasiparticle renormalization factor Z as a
function of $r_{s}$ obtained with different approximations in
comparison with QMC results.\cite{prb_50_1391,prl_107_110402} Also shown are the results based on
local field factors (LFF) which were copied from the
Ref.[\onlinecite{prb_77_035131}]. Abbreviations associated with QMC methods: BF - backflow; SJ - Slater-Jastrow; DMC - diffusion Monte Carlo; VMC - variational Monte Carlo; RMC - reptation Monte Carlo.} \label{Z_factor}
\end{figure}

\begin{figure}[t]
\centering
\includegraphics[width=8.0 cm]{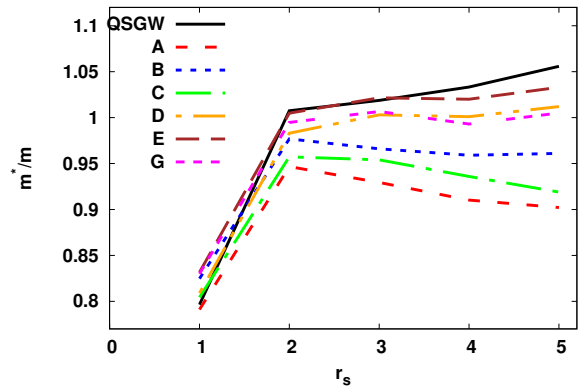}
\caption{(Color online) Effective mass of 3D electron gas as a
function of $r_{s}$. The results based on local field factors (LFF)
are taken from the Ref.[\onlinecite{prb_77_035131}].} \label{m_eff}
\end{figure}

Below we present quantities which show relatively slow convergence with respect to the quality of the momentum discretization mesh (q-mesh). They are the renormalization factor Z, the effective mass $m^{*}/m$, and the compressibility $\kappa$. Our present computational resources and specifics of our code didn't allow us to reduce an error bar on these quantities below 3-5\%. Still, we believe that the accuracy is good enough to make certain observations. The quasiparticle renormalization factor and the effective electron mass are presented in Figures \ref{Z_factor} and \ref{m_eff} correspondingly. In order to evaluate them, we used the following formulae (with $k_{F}$ being the Fermi momentum and $\mu$ being the chemical potential)
\begin{align}\label{zf}
Z=\Big\{1-\frac{\partial Im\Sigma(k,i\omega)}{\partial (i\omega)}|_{k=k_{F},\omega=\mu}\Big\}^{-1},
\end{align}
and
\begin{align}\label{ms}
\frac{m^{*}}{m}=\frac{Z^{-1}}{1+\frac{m}{k_{F}}\frac{\partial Re \Sigma(k,\omega)}{\partial k}|_{k=k_{F},\omega=\mu}},
\end{align}
correspondingly. We compare our results for Z-factor with available QMC data.\cite{prb_50_1391,prl_107_110402} Unfortunately, QMC data involve an extrapolation to the thermodynamic limit and show considerable variations for this quantity. We also compare our results with the ones based on local field factors\cite{prb_77_035131} (LFF) which include a certain amount of QMC input. Because of insufficient convergence of our results and the uncertainty in QMC data it is hard at this point to give a certain conclusion about which scheme provides the most accurate renormalization factor. If we assume that BF-VMC and BF-RMC results are the best, then we can state that our schemes D, E, and G are within uncertainty of QMC data. QSGW also shows good performance for this particular quantity.

It is interesting that the results for the effective mass ($r_{s}>2$) obtained from vertex corrected schemes lie in between the results obtained in scGW and QSGW approximations which serve as lower and upper limits correspondingly. One can also make an important observation that the $r_{s}$ dependence of the effective mass (for $r_{s}>2$) is weaker than similar dependence of the Z-factor suggesting that the frequency derivative and momentum derivative in Eq.(\ref{ms}) are canceling out considerably. This fact can have certain implications because there are theories (for example Dynamical Mean Field Theory) which stress frequency dependence but ignore momentum dependence of self energy in localized regime (low density).

\begin{figure}[t]
\centering
\includegraphics[width=8.0 cm]{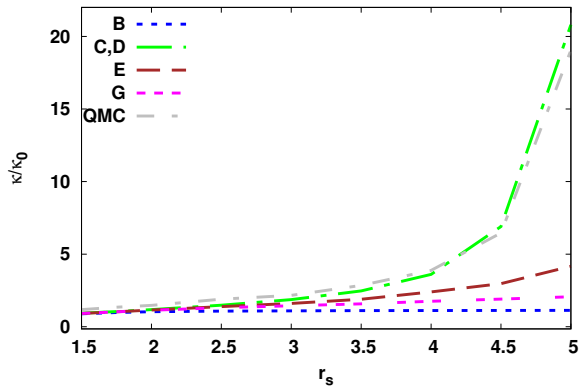}
\caption{(Color online) Compressibility of the electron gas. The QMC results have been obtained from the derivatives of the chemical potential with respect to the $r_{s}$. $\kappa_{0}$ is the compressibility of the non-interacting electron gas. To make the plot, we used the vertex $\Gamma_{\Theta}$ in schemes C, D, and E, and the vertex $\Gamma_{W}$ in scheme G.} \label{compress}
\end{figure}

We also checked, how well our calculated vertex functions reflect the presence of a pole in the compressibility (at about $r_{s}=5.25$, "dielectric catastrophe"\cite{prb_84_245134}). In Fig. \ref{compress} we compare our calculated compressibilities with the results based on the QMC data. We have obtained the QMC compressibility $\kappa$ as the derivative of the chemical potential $\mu$ ($=\frac{1}{2m}k_{f}^{2}+V_{xc}$) with respect to $r_{s}$: $\frac{1}{\kappa}=-\frac{1}{4\pi r_{s}^{2}}\frac{d\mu}{d r_{s}}$. In our vertex-corrected schemes, we have evaluated the compressibility from the ratio of two limits of vertex function ($\Gamma_{q}$ and $\Gamma_{\nu}$) and the effective mass $m^{*}/m$ (see for instance Ref.[\onlinecite{prb_84_245134}])

\begin{align}\label{compr}
\frac{\kappa}{\kappa_{0}}=\frac{m^{*}}{m}\frac{\Gamma_{q}}{\Gamma_{\nu}},
\end{align}
with two limits of vertex function defined in Appendix \ref{lim_det}.

It is clear from Figure \ref{compress} that the behavior consistent with the presence of a pole in the compressibility can only be obtained based on the vertex from schemes C and/or D. In other words, only the vertex corresponding to physical polarizability can be useful for compressibility evaluation. It is interesting, that additional self-consistency iterations for vertex function in schemes E and G (as compared to scheme D) which only slightly change the one-electron spectra, worsen significantly the quality of the calculated compressibility. Also, it is obvious that the first order vertex (scheme B) is totally insensitive to the presence of a pole in compressibility.

\section*{Conclusions}
\label{concl}

In conclusion we have applied the self-consistent diagrammatic
approaches based on the Hedin equations to study the properties of the
3D HEG. We have found, that the inclusion of the most important
diagrammatic sequences can reproduce the
one-electron spectra and dielectric properties of the HEG in the range of
metallic densities with good accuracy. For the one-electron spectra, the corrections to polarizability and to self
energy are equally important. For dielectric properties the
vertex correction to self energy is of secondary importance. In all cases, the important conclusion is that the calculation of polarizability should follow, as close as possible, its definition as a functional derivative of the density with respect to the total electric field. Our conclusions concerning one-electron spectra of 3D electron gas are similar to the conclusions made earlier for the spectra of alkali metals and semiconductors in Refs.[\onlinecite{prb_94_155101,prb_95_195120}], namely, that the best spectra are obtained when the set of diagrams for  polarizability is obtained from BSE, whereas the first order vertex correction is applied to the self energy (schemes D, E, and G). Our benchmarks quantified the inaccuracy the QSGW approximation to predict one-electron spectra of the electron gas at metallic densities (approximately 15\% error). We track this inaccuracy to the poor description of screening in QSGW approach (with an error up to a factor 2-3 in the physically important range of momenta). Concerning the use of the vertex-corrected schemes for the calculation of spectra, one can advocate scheme D, which combines good accuracy and computational efficiency (time-consuming BSE has to be solved only once).

\section*{Acknowledgments}
\label{acknow}

This work was   supported by the U.S. Department of Energy, Office of Science, Basic
Energy Sciences as a part of the Computational Materials Science Program. GK was supported by the
Simons Foundation under the Many Electron Problem collaboration.

\appendix

\section{Evaluation of two limits of the vertex function}\label{lim_det}

\begin{figure}[b]
\centering
\includegraphics[width=8.0 cm]{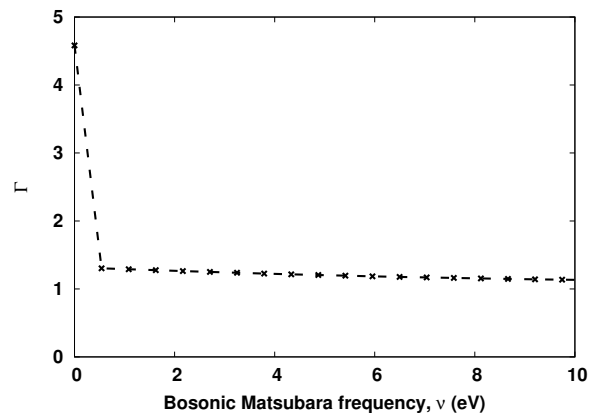}
\caption{Limiting behavior of the vertex function $\Gamma(k_{f},\pi/\beta;q=0,\nu)$ at small $\nu$ for $r_{s}=4$. Symbols '$\times$' show the calculated data points (discrete Matsubara's frequencies), line is drawn for convenience.} \label{vrt_plot}
\end{figure}

The two limits of the vertex function, $\Gamma_{q}$ and $\Gamma_{\nu}$ entering the Eq.(\ref{compr}), were defined as follows. In the momentum-frequency representation, the vertex function (solution of the Eq.(\ref{Vert_0})) can be conveniently considered as dependent on fermionic momentum+frequncy (\textbf{k},$\omega$) and on bosonic momentum+frequency (\textbf{q},$\nu$), i.e. $\Gamma$(\textbf{k},$\omega$;\textbf{q},$\nu$). In these variables, $\Gamma_{q}=\lim_{q\rightarrow 0} \Gamma(k=k_{F},\omega\rightarrow 0;q,\nu=0)$ and $\Gamma_{\nu}=\lim_{\nu\rightarrow 0} \Gamma(k=k_{F},\omega\rightarrow 0;q=0,\nu)$, with two vectors, \textbf{k} and \textbf{q}, assumed to be parallel. Quantities $\Gamma_{q}$ and $\Gamma_{\nu}$ are related to the quasiparticle renormalization factor $Z$ and the Landau Fermi liquid parameter $F^{s}_{0}$: $\Gamma_{\nu}=1/Z$, $\Gamma_{q}=\frac{1}{Z(1+F^{s}_{0})}$. In order to demonstrate that the above two limits are well defined numerically, we have plotted the vertex function $\Gamma_{\Theta}$ from scheme D in the Fig.\ref{vrt_plot}. The limit $\Gamma_{q}$ corresponds to the value of the function at $\nu=0$ exactly (approximately 4.58 on the graph). The limit $\Gamma_{\nu}$ corresponds to the extrapolation of the function to zero frequency (approximately 1.31 on the graph).


\end{document}